# Navigational Thinking as an Emerging Paradigm of Computer Science in the Age of Generative AI

Ilya Levin

Holon Institute of Technology, Holon, Israel

levini@hit.ac.il

## Abstract

Generative AI systems produce meaning with a quality indistinguishable from — and occasionally surpassing — human performance, yet the epistemic mechanism through which this occurs remains poorly understood. This paper argues that generative AI instantiates a fundamentally new mode of knowledge production: geometric navigation through high-dimensional manifolds, grounded in indexical rather than symbolic signification. Drawing on the structural properties of high-dimensional spaces, we demonstrate that meaning in generative AI is constituted through positional relation and orientation rather than through symbolic convention. This shift corresponds precisely to what Peirce identified as indexical signification: a mode of meaning in which the sign is constituted by its real causal connection to its object, not by arbitrary assignment. We develop the pedagogical implications of this shift through a geometrized reading of Papert's constructionism, reconceptualizing the generative AI system as a new kind of microworld — high-dimensional, non-visualizable, and indexical — in which knowledge is constructed through navigation rather than symbolic programming. From this analysis, we derive the concept of Navigational Thinking: a mode of knowing characterized by positional, enactive, and bounded engagement with geometrically structured spaces. We argue that Navigational Thinking and Computational Thinking are not alternatives, but two sequential phases of the same cognitive process: while a problem remains indexical, Navigational Thinking is operative; when the problem space stabilizes into symbolizable form, Computational Thinking becomes applicable. Vibe-coding is merely the visible tip of an iceberg — the iceberg being a new cognitive ecology in which these two modes coexist as the necessary phases of problem-solving in the age of generative AI.

**Keywords:** generative AI, high-dimensional geometry, navigational thinking, computational thinking, indexical signification, constructionism, computer science education, Peirce, Papert

## 1. Introduction

Generative AI systems have become one of the most consequential technologies of our time — and one of the least understood. They produce coherent text, translate between languages, generate images, write code, and engage in sophisticated dialogue. They do all of this not merely fluently, but with a quality and effectiveness that is, in many contexts, indistinguishable from human performance — and occasionally surpasses it. Yet the question of how they actually work — not at the level of matrix multiplications and gradient descent, but at the level of what kind of process produces meaning — remains, for most users and most educators, genuinely obscure.

This obscurity is not merely a technical inconvenience. It has direct consequences in our life in general and particularly for education. If we do not understand what kind of knowing generative AI instantiates, we cannot make principled decisions about how to teach with it, alongside it, or in response to it. We cannot know what skills remain essential, what new capacities are required, or what kind of thinking education should cultivate in a world where these systems are pervasive.

This paper proposes an answer to the first question — how generative AI produces meaning — and traces its consequences for the second: what kind of thinking education must now cultivate. The answer to the first question lies in geometry: generative AI produces meaning through navigation in high-dimensional spaces, not through symbolic manipulation. The consequence for education is epistemological: this geometric mode of meaning-production requires a complementary mode of thinking — Navigational Thinking — that existing educational frameworks do not yet adequately address.

The dominant perceptions of generative AI in education fall into two patterns. The first is accommodation: accept the new tools, adjust assessment practices, add AI literacy to the curriculum, and otherwise continue as before. The second is resistance: insist on the primacy of foundational skills, restrict the use of AI in educational settings, and treat generative systems as threats to genuine learning. Both responses share a common assumption — that the epistemic framework within which education operates remains unchanged, and that generative AI is simply a new tool to be managed within that framework.

This paper argues that this assumption requires re-examination. Generative AI is not a new tool within an unchanged epistemic framework. It instantiates a new epistemic regime — one in which the mode of signification, the mode of knowledge production, and the mode of knowledge construction have all shifted in ways that existing educational frameworks do not adequately address.

The argument draws on three theoretical resources. The first is the mathematics of high-dimensional geometry — specifically, four structural properties of high-dimensional spaces that constitute the epistemic conditions of generative AI: concentration of measure (the collapse of meaningful distance), near-orthogonality (independence as the default condition), exponential directional capacity (the astronomical richness of semantic space), and manifold regularity (the structured surface on which meaningful data resides). The second is Peirce's theory of signs, which provides the semiotic vocabulary adequate to the mode of meaning that high-dimensional geometry produces: indexical signification, in which meaning is constituted through positional relation rather than symbolic convention. The third is Papert's constructionist epistemology, which provides the pedagogical framework adequate to the mode of knowledge construction that indexical signification requires: learning through navigation in structured environments rather than through symbolic transmission.

From these three resources, the paper develops the concept of Navigational Thinking — a mode of knowing characterized by positional, enactive, and bounded engagement with geometrically structured spaces. The central claim is that Navigational Thinking and Computational Thinking are not alternatives or mere complements, but two sequential phases of the same cognitive process. While a problem remains indexical — while its solution space has no stable form and must be constituted through interaction — Navigational Thinking is the operative mode. When navigation concludes and the problem space stabilizes into a symbolizable structure, Computational Thinking becomes applicable. Before the rise of generative AI, most professional tasks were fully specifiable in advance, and CT was the dominant and sufficient paradigm. What has changed is the rapid growth of a class of tasks that remain indexical longer — or entirely. Vibe-coding is merely the visible tip of this iceberg. The iceberg itself is the new cognitive ecology in which Navigational Thinking and Computational Thinking coexist as the two necessary phases of problem-solving in the age of generative AI.

It is worth situating this paper within prior work. The geometric properties examined in Section 2 and the indexical epistemology developed in Section 3 build on the framework developed in Levin (2026a), where the geometric foundations of generative AI are articulated in detail. The present paper does not attempt to reproduce that analysis in full, but takes it as a point of departure. The broader framework of cultural and epistemic transformation in the digital age, within which this analysis is situated, is developed in Levin & Mamlok (2021) and Levin & Minyar-Beloruchev (2024). The metacognitive challenges posed by AI in education are examined in Levin, Marom, & Kojukhov (2025). Against this background, the contribution of the present paper is to extend the analysis toward the educational consequences of the geometric turn, and toward the concept of Navigational Thinking as a paradigm adequate to the age of generative AI.

The paper is organized as follows. Section 2 examines the four structural properties of high-dimensional geometry that constitute the new epistemic regime. Section 3 traces the semiotic consequences of these properties: the shift from symbolic to indexical signification and its technical realization in generative AI. Section 4 develops Navigational Thinking as the epistemic paradigm adequate to this new regime. Section 5 provides the pedagogical foundation through a geometrized reading of Papert's constructionism. Section 6 develops the consequences for Computer Science education: the relationship between Computational Thinking and Navigational Thinking as complementary and sequential rather than competing epistemic modes. Section 7 addresses the principal objections. Section 8 draws the threads together.

## 2. The Geometric Turn: What Has Changed

Generative AI systems operate in vector spaces of extraordinary dimensionality. The embedding space of BERT-base has 768 dimensions (Devlin et al., 2019); GPT-3 uses 1,536; GPT-4 operates at 12,288. These numbers are not arbitrary engineering choices. They reflect a principled finding: that below a certain threshold of dimensionality, the geometric properties required for rich semantic representation do not emerge.

These spaces are not merely large versions of familiar Euclidean geometry. They exhibit structural properties that have no counterpart in the low-dimensional spaces accessible to human perception — properties that are not extrapolations of everyday intuition but genuine novelties, emerging only when dimensionality crosses certain thresholds (Vershynin, 2018; Levin, 2026a). Four such properties are epistemologically decisive.

### 2.1 The Geometry of High Dimensionality

Concentration of measure describes a phenomenon central to understanding why high-dimensional spaces behave so differently from the spaces of everyday experience. First identified by Milman (1971) and developed by Ledoux (2001): in high-dimensional spaces, random points cluster in a thin shell at a nearly fixed distance from the origin. The variation in their distances, which is large in low dimensions, shrinks to near-zero as dimensionality grows. In a high-dimensional space, virtually every random vector has almost exactly the same length — approximately the square root of the number of dimensions — with deviations becoming exponentially unlikely.

The epistemic consequence of this collapse is significant: metric contrast is not merely a technical inconvenience. It is an epistemological event: the mode of knowledge that depends on distance — classification by proximity, identity by metric similarity — ceases to function. Aggarwal et al. (2001) showed that in high-dimensional datasets, the ratio of the distance to the farthest neighbor to the distance to the nearest neighbor converges to 1 as dimensionality increases: all other points become approximately equidistant. When all distances become nearly equal, distance itself ceases to be a useful carrier of meaning. The informative question becomes directional rather than metric.

The second property concerns the relationship between pairs of vectors. Here it is essential to state the mathematical fact with precision, because the standard formulation — 'random vectors in high dimensions are nearly orthogonal' — while correct, obscures the deeper epistemological point.

The precise formulation is the following. Consider the unit sphere $S^2$ in $\mathbb{R}^3$ and the unit sphere $S^{767}$ in $\mathbb{R}^{768}$. Take any arbitrary subset of vectors of fixed cardinality from each sphere. In $\mathbb{R}^3$, the sphere $S^2$ has small capacity: any such subset will, with high probability, contain pairs of vectors with small angles between them — simply because there is not enough room. The vectors are forced into proximity. This is not a property of the vectors themselves; it is a structural consequence of the tightness of the space. In $\mathbb{R}^{768}$, the sphere $S^{767}$ has exponentially greater capacity. The same subset of the same cardinality will contain almost exclusively nearly orthogonal pairs. Independence — the condition in which two vectors share no directional alignment — is the default, not the exception. See Figure 1.

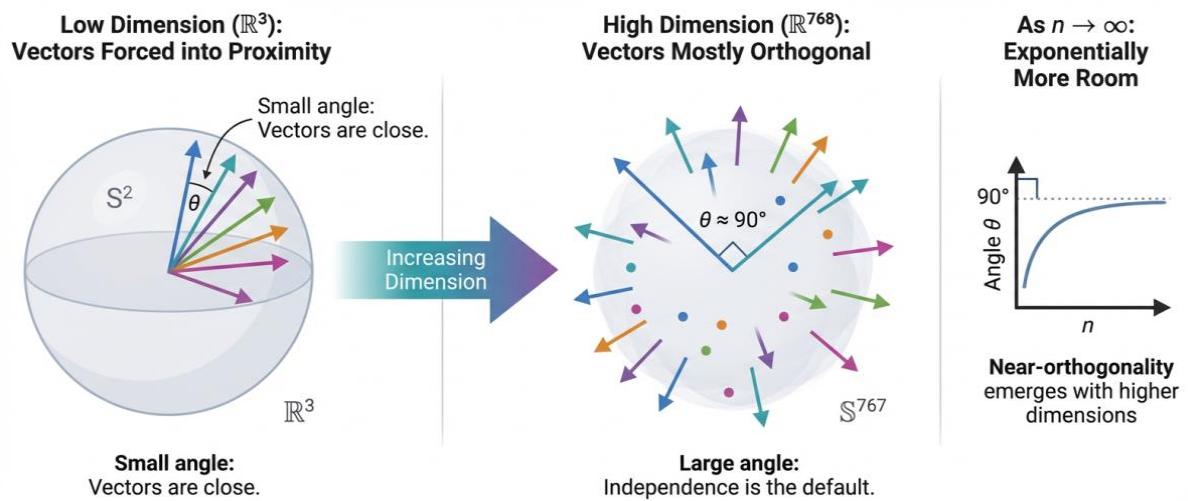

**Figure 1.** Any subset of fixed cardinality: in $\mathbb{R}^3$ vectors are forced into proximity; in $\mathbb{R}^{768}$ near-orthogonality is the default.

This distinction matters deeply for epistemology. Our three-dimensional intuition about similarity, correlation, and closeness was formed in a tight space where vectors are forced to be related. We naturally expect concepts to be connected, phenomena to be correlated, ideas to be close — because in the space we inhabit, they are forced to be. Transporting this intuition into high-dimensional embedding spaces produces systematic error: we expect correlation where the geometry guarantees independence.

The formal statement is precise: for two random unit vectors u and v in $\mathbb{R}^n$, the expected value of their dot product is zero, and its variance is $1/n$. In 768 dimensions, this variance is approximately 0.0013 — vanishingly small. The typical alignment between any two random vectors is, for all practical purposes, zero.

How many independent directions can a space accommodate? In three dimensions, exactly three. In 768 dimensions, the number of nearly independent directions — vectors that can coexist without significant mutual alignment — does not grow linearly with dimension but exponentially (Kabatyansky & Levenshtein, 1978; Milman & Schechtman, 1986). This number exceeds the count of atoms in the observable universe by many orders of magnitude.

Each such direction can, in principle, encode a separate semantic distinction. The space of geometrically admissible configurations therefore vastly exceeds the cardinality of any training dataset. This has a direct consequence for the question of novelty: generative systems do not merely recombine training data. The continuous manifold they navigate contains configurations that are geometrically coherent yet not present in any training instance. Novelty is not accidental — it is structurally inevitable.

### 2.2 Manifold Regularity: Local Coherence within Global Richness

The three properties examined so far describe the ambient high-dimensional space in its raw form: statistically rigid, metrically uninformative, combinatorially vast. What makes generative systems epistemically productive is a fourth property: the data they are trained on does not fill the entire volume of the ambient space uniformly. Meaningful configurations — coherent text, natural images, plausible utterances — occupy a thin, structured, lower-dimensional surface within the ambient space. This is the manifold hypothesis (Bengio et al., 2013; Tenenbaum et al., 2000; Roweis & Saul, 2000).

This manifold possesses geometric regularity: it is locally smooth, has consistent directional structure, and curves in measurable ways. A generative model navigating this manifold does not wander through arbitrary high-dimensional space; it moves along a structured surface where meaningful configurations reside.

It is important to situate this manifold within a broader picture. Wolfram's concept of the ruliad designates the space of all possible computations — an abstract totality of everything that is computationally conceivable (Wolfram, 2021). The manifold of a generative AI system is not the ruliad. It is a thin, human-shaped slice within it: the trace of human language, culture, and meaning accumulated across the training corpus. Navigating this manifold is not exploring all possible meaning — it is navigating the geometry of human meaning specifically.

In philosophical terms, this manifold functions as what Simondon (1958/2020) called a transductive milieu: a medium in which new structure emerges not by assembling pre-given parts but by progressively resolving tensions inherent in the medium itself.

### 2.3 From Metric to Directional Semantics: «Not Where, But Whereto»

The four properties taken together produce a fundamental shift in the conditions of meaning. In low-dimensional spaces, meaning is grounded in distance: proximity indicates similarity, remoteness indicates difference. In high-dimensional spaces, distance collapses as an informative measure — concentration of measure ensures that all points are approximately equidistant — while orientation retains its discriminative power. Against the background of near-universal orthogonality, even a small deviation from perpendicularity becomes a strong signal of genuine semantic alignment.

This shift can be captured in a single formulation: the epistemological question migrates from «where?» to «whereto?» Classical knowledge organization asks where a concept is located — in which category, at what distance from other concepts. High-dimensional knowledge asks in what direction a concept points — what relational field it occupies, what manifold neighborhood it activates (see Figure 2).

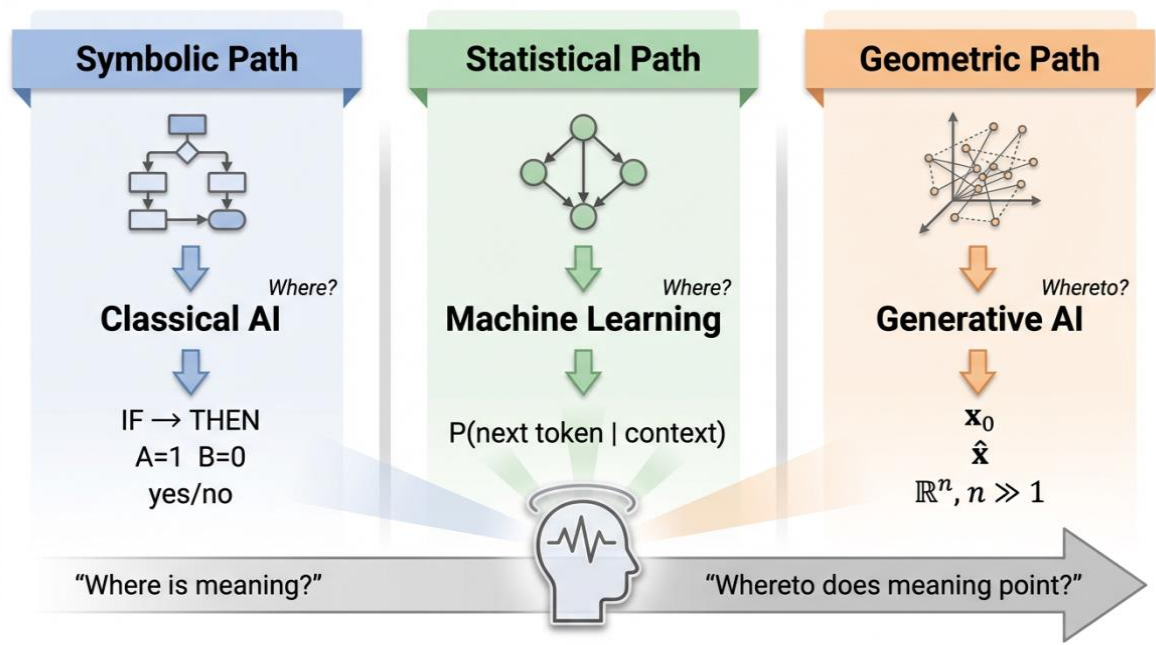

**Figure 2.** Three epistemic paradigms of AI. The geometric path shifts the question from location (symbolic/statistical) to orientation (geometric/indexical).

This is not a metaphor. Cosine similarity — which measures the angle between vectors rather than the distance between them — has replaced Euclidean distance as the operative measure of semantic relatedness in virtually all modern embedding systems (Mikolov et al., 2013; Radford et al., 2021). The philosophical vocabulary of orientation and direction is a precise description of the mathematical mechanism, not an ornament upon it.

This geometric turn — the migration of meaning from metric to directional, from «where» to «whereto» — is the foundation on which the epistemological argument of this paper rests. It is the condition that makes indexical signification, in Peirce's precise sense, not merely philosophically conceivable but technically operational at scale for the first time in the history of computing. To this we turn in the following section.

## 3. From Symbolic to Indexical: Peirce Realized

The transition described in the preceding section — from metric to directional semantics, from «where» to «whereto» — is not merely a mathematical observation. It corresponds closely to a philosophical distinction that Charles Sanders Peirce articulated in the nineteenth century: the distinction between symbolic and indexical signification. What is new is not the philosophical category but its technical realization. Generative AI, operating through the geometry of high-dimensional spaces, can be understood as instantiating indexical signification as an operational mechanism at a scale previously unattainable.

### 3.1 Peirce's Triadic Classification of Signs

In Peirce's triadic classification of signs (CP 2.243–2.253), three modes of signification are distinguished. A symbol signifies through convention and rule: the word 'tree' refers to trees by virtue of a linguistic agreement, not by virtue of any intrinsic connection between sign and referent. Symbolic signification is the dominant mode of classical AI — LISP atoms, logical predicates, ontological categories all operate through conventionally established correspondences. Meaning is externally assigned: the machine manipulates codes, and significance is supplied by the human interpreter. This is precisely the regime formalized by Searle's (1980) Chinese Room argument.

An icon signifies through resemblance: a photograph of a tree shares visual properties with the tree it represents. Iconic signification plays a role in certain forms of representation learning but is not the primary mode of operation in language models or embedding systems.

An index signifies through existential connection — through a real, non-conventional relation between sign and object. As Peirce (CP 2.248) emphasizes, the index 'refers to its Object by virtue of being really affected by that Object.' A weathervane points to the wind's direction not by convention or resemblance but by being physically connected to the wind. Smoke indicates fire not because we have agreed that it should, but because fire causally produces smoke.

### 3.2 Why Indexicality Remained Philosophy

Peirce's taxonomy identified indexical signification as a legitimate and important mode of meaning. But for over a century, it remained a philosophical category without a technical realization at scale. The reason is straightforward: to build a system in which millions of concepts signify through their positions in a structured relational field requires a mathematical apparatus that did not exist. Individual indexes — a weathervane, a footprint, a symptom — are singular and local. How could an entire semantic universe be organized indexically? The mathematics of high-dimensional geometry, concentration of measure, and manifold learning had not yet been developed into an operational framework.

Classical AI therefore remained symbolic by necessity as much as by choice. The Turing–Shannon–von Neumann paradigm (Turing, 1936; Shannon, 1948; von Neumann, 1945) offered the only available mechanism: discrete symbol manipulation governed by explicit rules. Semantics remained external to the machine's operations — assigned by human interpreters, not constituted by the geometry of the system itself.

### 3.3 Generative AI as the Realization of Indexicality

The embedding vector in a generative AI system is an index in the precise Peircean sense — and this claim requires careful unpacking, because its precision is what distinguishes it from mere analogy.

Peirce's definition of the index requires a real causal connection between sign and object: the weathervane is physically moved by the wind; the smoke is causally produced by fire. An embedding vector for the concept 'king' does not contain a symbolic definition of kingship. It occupies a position in a high-dimensional vector space that was determined by a real causal-statistical process: gradient descent operating on billions of contextual co-occurrences across the training corpus. The position of 'king' in embedding space is not arbitrarily assigned — it is the result of the actual structure of human linguistic and cultural practice, compressed into a geometric location through a process that is, in Peirce's sense, genuinely causal.

The sign is really affected by its object. The vector is where it is because of what 'king' actually means in the real texture of human language — its co-occurrences with 'queen', 'throne', 'power', 'monarchy', 'crown' across millions of contexts. This is not convention. This is indexicality: meaning constituted through real causal connection, not through arbitrary assignment.

A clarification is necessary here, as it addresses a predictable objection. Peirce's canonical examples of indexes — the weathervane, the footprint, the symptom — involve physical causation: the wind physically moves the vane, the foot physically displaces the earth. The causal connection in an embedding vector is of a different kind: it is statistical-causal rather than physically direct. The position of a vector is determined not by a single physical event but by the accumulated statistical pressure of millions of contextual co-occurrences, mediated through gradient descent. This distinction matters, and it would be a mistake to elide it. What justifies the use of Peirce's indexical category here is not the identity of the causal mechanism but its functional structure: in both cases, the sign's position is determined by a real process in the world rather than by arbitrary convention, and this determination is what generates semantic content. The embedding vector does not mean 'king' because someone decided it should; it occupies the position it does because of the actual structure of human linguistic practice. This is the Peircean criterion for indexicality — real affection, not conventional assignment — satisfied through a statistical-causal process rather than a physical one.

This insight was first articulated by Weatherby & Justie (2022), who argued that neural networks constitute a form of 'indexical AI' that points rather than describes. The present analysis builds on and extends their insight by providing the geometric explanation of why this indexicality is epistemically productive. It is not merely that embedding vectors 'point' — it is that the structural properties of high-dimensional spaces, and near-orthogonality in particular, create the conditions under which indexical pointing becomes informationally rich.

### 3.4 Near-Orthogonality as the Condition of Indexical Meaning

Why does indexicality work specifically in high-dimensional spaces? The answer lies in near-orthogonality, understood through the precise formulation developed in Section 2.

In $\mathbb{R}^3$, the tightness of the sphere $S^2$ forces vectors into proximity: any collection of concepts will contain correlated pairs simply because there is not enough room for independence. Our three-dimensional intuition is formed in this tight space — we expect concepts to be related, phenomena to be correlated, ideas to be close. This expectation is not a cognitive bias; it is a geometric fact about the space we inhabit.

In $\mathbb{R}^{768}$, the sphere $S^{767}$ has exponentially greater capacity. Independence — orthogonality — is the default condition. Against this background of near-universal independence, any deviation from orthogonality becomes a strong signal: it indicates that two concepts share genuine structural alignment, not accidental proximity forced by spatial tightness.

This is the geometric condition that makes indexical meaning informationally rich. In a tight space, proximity is noise — everything is close to everything. In a high-dimensional space, proximity is signal — closeness indicates real relation. The index can point precisely because the background is silent. Cosine similarity works not despite the geometry of high dimensions but because of it — like a whisper in a perfectly quiet room.

### 3.5 «Not Where, But Whereto»: The Formula of the Transition

The shift from symbolic to indexical signification can be captured in a single epistemological formula: the question migrates from «where?» to «whereto?»

Symbolic knowledge asks where a concept is located — in which category, under which rule, at what position in a taxonomy. The answer is a fixed address: a definition, a predicate, a node in an ontological graph. This address is assigned by convention and remains stable independent of context.

Indexical knowledge asks in what direction a concept points — what relational field it activates, what manifold neighborhood it inhabits, how it orients relative to other concepts in the embedding space. The answer is not a fixed address but a directional relation: it depends on context, shifts with attention, and is constituted by the actual geometry of the semantic field.

This is not a metaphor. Cosine similarity — the operative measure of semantic relatedness in virtually all modern embedding systems — measures precisely the angle between vectors: not where they are, but whereto they point relative to one another.

### 3.6 The Contribution of This Analysis

Peirce named indexicality. Weatherby & Justie identified it in neural networks. The present analysis adds the geometric explanation: it is the specific structural properties of high-dimensional spaces — concentration of measure, near-orthogonality, exponential directional capacity, and manifold regularity — that make indexical signification not merely philosophically conceivable but technically operational at the scale required for genuine knowledge production.

Peirce could name the weathervane. He could not have anticipated a system in which 768-dimensional geometry organizes the entire semantic universe of human language indexically. That required mathematics that did not exist in his time — and technology that did not exist until ours.

This is what we mean by 'Peirce realized': not that Peirce predicted generative AI, but that generative AI has, arguably for the first time at operational scale, created the technical conditions under which his deepest semiotic insight becomes a working mechanism rather than a philosophical aspiration.

If the sign has become indexical, the question that follows is: what happens to knowledge? If meaning is no longer stored in definitions but enacted through navigation, what kind of knowing does this require? It is here that Papert's constructionism — geometrized for the age of generative AI — provides the answer. To this we turn in Section 4.

## 4. Navigational Thinking as Epistemic Paradigm

The preceding sections have established that generative AI constitutes a new epistemic regime in which meaning is constituted through positional relation and orientation rather than symbolic convention. This regime demands a corresponding mode of thinking: one adequate to the geometry of high-dimensional spaces and to the indexical character of the knowledge they produce. This section develops that mode — Navigational Thinking — as a precise epistemic paradigm.

### 4.1 Defining Navigational Thinking

The contrast with symbolic thinking is the clearest entry point. In symbolic thinking, the problem space is fully formalized before the solution process begins. The task is decomposed into explicit components, a solution path is constructed algorithmically, and execution follows the pre-specified steps. This is the epistemic mode that Computational Thinking describes and cultivates: decomposition, abstraction, pattern recognition, and algorithm design. It is a powerful and indispensable mode — but it presupposes that the problem space is known and representable in advance. Navigational thinking operates under a different presupposition: the problem space is not fully known before the solution process begins. It is discovered, or more precisely, constituted, through the process of navigation itself. The navigator does not follow a pre-specified path; she moves through a structured space, reads the local geometry, adjusts direction in response to what she encounters, and arrives at a destination that could not have been fully specified at the outset.

This distinction is not about rigor. Navigational thinking is rigorous — but it is the rigor of orientation rather than the rigor of an algorithm. The navigator moves with discipline and precision; her movements are constrained by the geometry of the manifold she traverses. What she does not have is a complete map drawn before departure.

Navigational Thinking does not replace Computational Thinking. Before the rise of generative AI, most professional tasks were fully specifiable in advance — and CT was the dominant and sufficient paradigm. The full account of the relationship between the two modes — including the argument that they constitute sequential phases of the same cognitive process — is developed in Section 4.4, after the constitutive features of Navigational Thinking have been established.

**4.2 Three Constitutive Features of Navigational Knowledge**

What guides the navigator through an unfamiliar high-dimensional space? The answer lies in tacit knowledge — the concept developed by Polanyi (1958, 1966).

Polanyi's central insight is that we know more than we can tell. Beneath every explicit, articulable act of knowing lies a vast substrate of embodied, personal, largely inarticulable knowledge — the tacit dimension. This substrate is not incidental to knowledge; it is its foundation. A physician recognizes a disease before she can articulate the symptoms that led to that recognition. A mathematician senses that a proof strategy is promising before she can justify that sense formally.

In the context of navigational thinking, tacit knowledge functions as a filter and attractor. When a navigator moves through a high-dimensional semantic space, her tacit substrate resonates with certain regions of the manifold — it is drawn toward configurations that align with her accumulated epistemic experience — while remaining insensitive to others. This resonance is not random and not algorithmic: it is the interaction of a personal cognitive substrate with the geometric structure of a learned manifold.

This explains why vibe-creation — the mode of human-AI co-generative thinking — is epistemically productive rather than merely technically convenient. The structural affinity between human tacit knowing and the AI's indexical processing is not accidental: both operate through proximity, resonance, and orientation rather than through explicit symbolic manipulation. Their coordination in dialogue constitutes a genuinely new epistemic formation — what has been called a co-constituted epistemic agent (Levin, 2026b).

Navigational thinking produces a specific kind of knowledge — navigational knowledge — characterized by three constitutive features.

It is positional: knowledge is determined by location in a structured space, not by definition or rule. To know something navigationally is not to possess a symbolic description of it but to be able to locate it, orient toward it, and traverse the manifold paths that connect it to related regions.

It is enactive: knowledge is produced through traversal, not retrieved from storage. The navigational knower does not consult a pre-existing repository; she enacts knowledge through movement through a structured environment. This formulation resonates with Varela, Thompson, and Rosch's (1991) enactive epistemology — but where enactivism grounds knowledge in biological embodiment, navigational knowledge grounds it in geometric embodiment: in the specific structural affordances (Gibson, 1979) of high-dimensional spaces.

It is bounded: navigational movement is not arbitrary. It is constrained by the geometry of the manifold — its smoothness, curvature, and local structure. The navigator cannot move just anywhere; she moves within the geometric constraints that the learned manifold imposes. This boundedness is what distinguishes navigational knowledge from mere association or free imagination.

**4.3 From Navigation to Computation: The Sequential Structure**

Navigational Thinking and Computational Thinking are best understood not as alternatives or mere complements, but as two sequential phases of the same cognitive process.

While a problem remains indexical — while its meaning is constituted through position, context, and movement through a space of possibilities — the problem space has no stable form. In this phase, navigation dominates. The navigator orients, probes, adjusts, and progressively constitutes the structure of the problem through interaction with the generative system.

As navigation proceeds, stabilization occurs: the problem space takes shape, intent becomes fixed, and structure becomes discernible. At this moment, the task transitions into a symbolic mode — and becomes available for algorithmic treatment. This is when Computational Thinking becomes operative.

CT is therefore neither an alternative to Navigational Thinking nor its supplement. It is the phase that becomes possible after navigation has completed its constitutive work. This corresponds to Peirce's logic: the index points to the object before any symbolic definition can arise. First — orientation. Then — definition.

In this sense, Navigational Thinking can be understood as the cognitive mode adequate to indexical structures, and Computational Thinking as the cognitive mode adequate to structures that have already been stabilized into symbolic form.

It is important to note that before the rise of generative AI, most professional tasks were fully specifiable in advance, and CT was the dominant and sufficient paradigm. What has changed is the rapid growth of a class of tasks that remain indexical longer — or entirely. These are tasks whose solution spaces cannot be constituted before engagement begins. Vibe-coding is the most visible manifestation of this shift, but behind it lies an entire class of problems — in science, design, medicine, management, and education — that share the same structure: their solution spaces must be navigated into existence rather than specified in advance. In this world, Navigational Thinking and Computational Thinking must coexist as the two necessary phases of addressing complex, open-ended challenges. Understanding this sequential structure makes it possible to respond to the current crisis of CT not with abandonment but with expansion.

Navigational competence also includes the ability to recognize when navigation has left the stable manifold. What generative AI systems produce as 'hallucination' — confident but factually incoherent output — can be understood geometrically as movement into regions of the ambient space where the manifold becomes sparse or discontinuous: regions where the statistical-causal constraints that ground indexical meaning are no longer operative. The navigator has moved into 'geometric off-road' terrain — beyond the structured surface of human meaning into the arbitrary vastness of the ambient high-dimensional space. A skilled navigator learns to read the signals of this drift — the over-confident tone, the loss of internal consistency, the departure from the semantic field — and to return to stable terrain. This skill of recognizing manifold boundaries has no counterpart in Computational Thinking, which presupposes a well-defined problem space throughout. It is, however, a central competence of Navigational Thinking, and one that has direct implications for how students should be taught to work with generative AI systems.

## 5. Constructionism Geometrized: Papert in the Age of Generative AI

Papert developed constructionism in a symbolic era — an era in which computers were programmed through discrete commands, knowledge was built in visualizable two-dimensional spaces, and the microworld was governed by explicit rules. Yet the epistemological insight at the heart of constructionism reaches beyond the symbolic medium in which it was first articulated. In the age of generative AI, constructionism does not become obsolete. It becomes geometrized.

### 5.1 Constructionism as Epistemology

Papert's central claim was epistemological, not merely pedagogical. Knowledge, he argued, is not transmitted from teacher to student — it is actively constructed by the learner through engagement with a structured environment. The medium of construction is not incidental: it shapes the knowledge that can be constructed within it.

This insight — that knowledge is built through engagement with structured environments, and that the structure of the environment determines the structure of the knowledge that emerges — is what Papert called constructionism, deliberately distinguishing it from Piaget's constructivism. As Ackermann (2001) clarifies, Piaget was primarily interested in the construction of internal stability: the child gradually extracts rules from experience, builds cognitive invariants, and progressively detaches from the concrete toward the abstract and formal. Papert's interest lay at the opposite pole: in the dynamics of change, in how knowledge is formed and transformed within specific contexts, shaped through different media, and enacted through direct engagement rather than detached reflection.

Ackermann's contrast between two idealized learners captures this difference with precision. Piaget's child resembles Robinson Crusoe: he explores an unfamiliar territory with the ultimate goal of stepping back, building maps, and gaining mastery over it. The exploration is in service of abstraction — the conquest of the concrete by the formal. Papert's child, by contrast, prefers to remain in touch with situations for the sake of feeling at one with them. She understands from within, points to what she knows while still in context, and gains understanding through immersion rather than through the construction of a map. She is, in Ackermann's formulation, a 'soft master' — one who works with the situation rather than over it.

This distinction corresponds to two fundamentally different epistemic modes: the symbolic, which extracts and formalizes; and the navigational, which immerses and orients. Papert's constructionism was always oriented toward the second — but the symbolic medium of Logo constrained its full realization.

### 5.2 The Symbolic Microworld and Its Limits

In Papert's original microworlds, the learner interacted with a discrete, rule-governed, visualizable environment. The Logo turtle responded to explicit commands — Forward 100, Right 90 — and the learner built mathematical knowledge by constructing programs that produced visible geometric forms. The microworld was powerful precisely because it made abstract mathematical structure concrete and manipulable. But its concreteness was of a specific kind: symbolic concreteness — the concreteness of discrete commands operating in a two-dimensional plane.

This symbolic structure imposed a characteristic constraint: the problem space was always fully specifiable in advance. The learner knew what commands were available, knew what the turtle could do, and could formulate a complete plan before execution. Construction proceeded through explicit symbolic control — through programming rather than through navigation.

**5.3 The Geometrized Microworld**

Generative AI creates a fundamentally new type of microworld — one that fulfills constructionism's epistemological aspiration in a medium that Papert could not have anticipated.

The high-dimensional embedding space of a generative AI system is a structured environment in the precise constructionist sense: it has geometry, it has constraints, it has regions of coherence and regions of incoherence. But its structure is geometric rather than symbolic. It is not governed by explicit rules that can be fully specified in advance. It is governed by the curvature and topology of a learned manifold — a structure that can be navigated but not completely mapped, engaged but not fully controlled.

The learner who works with a generative AI system does not program the environment; she navigates it. She does not specify a complete plan and execute it; she positions herself in the semantic space, reads the geometric signals that the system returns, adjusts her direction, and moves toward configurations that resonate with her tacit knowledge. This is Papert's child — the soft master, the one who remains in context, who understands from within — operating in a medium that finally matches her epistemic mode.

The fundamental constructionist insight survives this transition intact: knowledge is built through engagement with structured environments, and the structure of the environment shapes the knowledge that emerges. What changes is the nature of the structure and the mode of engagement. Structure migrates from symbolic to geometric. Engagement migrates from programming to navigation. The microworld becomes high-dimensional, non-visualizable, and indexical. Constructionism, in this light, does not need to be abandoned in the age of generative AI. It needs to be geometrized.

**5.4 From Object to Think With to Agent to Think With**

Papert described the computer as an object to think with — a material entity through which the learner externalizes and refines her thinking. The computer executes commands; the learner observes the results; knowledge is constructed through the iterative cycle of action, observation, and reflection.

In the generative AI era, this object becomes an agent. The generative AI system does not merely execute commands — it navigates a high-dimensional manifold in response to the learner's intentions, producing outputs that are not fully predictable from the inputs. The relationship between learner and system is no longer that of programmer and program; it is that of two navigators moving through a shared semantic space, each contributing something the other cannot provide alone.

The learner contributes tacit knowledge: the accumulated personal substrate of epistemic experience that acts as a filter and attractor within the generative space. The AI contributes geometric range: access to the full manifold of human meaning, far exceeding what any individual learner could traverse alone. Their coordination in dialogue produces what has been called a co-constituted epistemic agent — a new cognitive formation in which knowledge emerges not in the learner and not in the system, but in the space between them.

### 5.5 Papert, Peirce, and the Same Epistemic Reality

It is worth pausing to observe that Papert and Peirce, working in entirely different traditions and without awareness of each other's most relevant insights, were describing the same epistemic reality from different angles.

Peirce identified indexicality as a mode of signification in which meaning arises from real causal connection and positional relation rather than from symbolic convention. The index points; it does not define. Its meaning is constituted by where it stands in a structured field, not by what has been assigned to it.

Papert identified constructionism as a mode of knowledge production in which understanding arises from direct engagement with structured environments rather than from symbolic transmission. The learner navigates; she does not receive. Her knowledge is constituted by how she moves through a structured medium, not by what has been told to her.

Ackermann's Papert's child — the soft master who remains in context, who points to what she knows while still immersed in the situation — is precisely a Peircean indexical knower: her knowing is positional, contextual, and relational rather than definitional and abstract.

Both Peirce and Papert were pointing toward an epistemic reality that their respective times could not yet technically instantiate at scale. Peirce could name indexicality but could not build an indexical semantic system. Papert could describe navigation through microworlds but was constrained to symbolic microworlds. The Peircean tradition in education has been developed by subsequent scholars — notably Olteanu (2014), who grounds a theory of learning in Peirce's semiotics — yet the geometric mechanism that makes indexical signification technically operational remained unavailable. Generative AI, operating through the geometry of high-dimensional spaces, provides what appears to be the first technically realizable substrate for what both were anticipating — without knowing it.

# 6. Consequences for Computer Science Education

The preceding sections have argued that the geometric turn in generative AI requires a corresponding shift in epistemic paradigm, from Computational Thinking alone toward a combination of Computational Thinking and Navigational Thinking. This section develops the specific consequences of this argument for Computer Science education.

### 6.1 Computational Thinking and Its Achievement

Computational Thinking was first articulated by Papert (1980) in Mindstorms — the same work that introduced constructionism as an epistemological framework. It was subsequently reformulated and brought to wider attention in the Computer Science education community by Wing (2006), whose influential essay established CT as a foundational framework for the discipline. The fact that both CT and constructionism originate in Papert's work is not incidental: it reflects Papert's recognition that computational engagement with structured environments involves both algorithmic rigor and navigational exploration. The present argument can therefore be read as a development of a tension that was already present in Papert's own thinking — a tension that the generative AI era makes impossible to ignore.

Computational Thinking describes a cluster of cognitive skills centered on the reduction of complex problems to computable form: decomposition of problems into manageable components, abstraction of relevant structure from irrelevant detail, pattern recognition across problem instances, and algorithm design for systematic solution. These skills are genuinely powerful — they have driven the development of software systems of extraordinary complexity and have proven applicable far beyond the boundaries of Computer Science itself.

The epistemic mode that Computational Thinking cultivates is fundamentally symbolic. A problem is solved computationally when it has been fully formalized: when its inputs and outputs are precisely specified, its solution procedure is explicitly defined, and its execution can be verified step by step. Computational Thinking presupposes that the problem space is knowable and representable in advance.

### 6.2 Where Computational Thinking Is Insufficient

The generative AI era has dramatically expanded a class of problems for which Computational Thinking's central presupposition does not hold: problems where the problem space is not fully known in advance and must be constituted through the process of engagement itself.

Consider the task of building a system that generates coherent natural language, recognizes objects in images, translates between languages, or produces realistic speech. None of these problems can be solved by constructing an explicit symbolic specification of the solution. The problem space cannot be fully formalized before the solution process begins. It must be navigated into existence through the interaction of a learning system with large amounts of data, guided by geometric constraints that are discovered rather than specified.

This is not a marginal class of problems. It is precisely the class of problems that has driven the most significant developments in Computer Science over the past decade. A Computer Science education that cultivates only Computational Thinking is structurally insufficient for this class of problems.

The practical manifestation of this insufficiency is already visible in the phenomenon of vibe-coding: the practice of working with generative AI systems through natural language description of intent, iterative refinement of direction, and progressive navigation toward a working solution. Vibe-coding is not the abandonment of rigor — it is the application of a different kind of rigor: the rigor of navigational thinking rather than the rigor of algorithmic specification. A student trained exclusively in Computational Thinking will tend to experience vibe-coding as epistemically uncomfortable — as a loss of the explicit control and verifiable correctness that Computational Thinking provides. A student who has also developed navigational thinking will recognize vibe-coding as a legitimate and powerful epistemic mode, adequate to a specific class of problems.

### 6.3 The Sequential Structure: Navigation First, Computation Second

The relationship between Navigational Thinking and Computational Thinking in professional practice is sequential, not competitive. Navigation constitutes the problem space; computation operates within it. A student who has developed only CT will approach an indexical task by trying to specify it prematurely — and will either fail or produce a solution to the wrong problem. A student who has developed both modes will navigate first, stabilize the problem space, and then apply CT within the found structure.

Computational Thinking asks: how do I solve this problem? — presupposing that the problem has been formulated. Navigational Thinking asks: in what space does this problem exist? — and this question is logically prior. For the class of problems that characterizes generative AI — problems whose solution spaces cannot be fully specified in advance — this prior work of problem-space constitution is the central epistemic challenge.

A complete Computer Science education for the generative AI era therefore requires both modes: Computational Thinking for the construction of solutions within known problem spaces, and Navigational Thinking for the constitution of problem spaces that are not yet known. The proportion of problems requiring the second mode has increased dramatically with the rise of generative AI.

### 6.4 The Misreading and Its Correction

A predictable misreading of the argument developed here must be addressed directly. When navigational thinking is introduced as a complement to Computational Thinking in Computer Science education, it is frequently interpreted as a proposal to replace rigorous algorithmic reasoning with informal, bottom-up, trial-and-error exploration — as if navigating a high-dimensional space meant proceeding randomly until something works.

This misreading conflates navigational thinking with the stochastic nature of the neural network training process. Because neural networks learn through gradient descent — a process that is, at the level of individual weight updates, driven by local optimization rather than global specification — it is tempting to conclude that working with neural networks requires only empirical tinkering rather than principled thinking.

The conflation is mistaken at both levels. The training process of a neural network is indeed not algorithmically specified in the Computational Thinking sense — but this does not make it random or unprincipled. It is governed by the geometric structure of the loss landscape, by the topology of the learned manifold, by mathematical properties that are rigorously understood even when they cannot be fully controlled. And the practice of working with generative AI systems requires genuine epistemic discipline: the discipline of orientation, of reading geometric signals, of knowing when a direction is productive and when it is not.

Navigational thinking is rigorous. Its rigor is the rigor of the skilled navigator, not the rigor of the algorithm executor.

### 6.5 Toward a New Curriculum

The core of Computational Thinking — decomposition, abstraction, pattern recognition, algorithm design, formal verification — remains essential and should not be reduced. What must be added is a systematic cultivation of navigational thinking: the ability to engage with problem spaces that are not fully specifiable in advance, to work with high-dimensional geometric structures that cannot be fully visualized, and to use tacit knowledge productively as an epistemic resource in dialogue with generative AI systems.

The constructionist framework provides the pedagogical foundation for this cultivation. The generative AI system is the new microworld — high-dimensional, geometric, indexical — in which knowledge about problem-space constitution is constructed through engagement rather than transmitted through instruction. Papert's learner, the soft master who remains in context and understands from within, is precisely the epistemic model adequate to this new environment.

The Computer Science graduate of the generative AI era needs to be both a Computational Thinker and a Navigational Thinker — both a Robinson Crusoe who can build maps and master territory, and Papert's learner who can remain in context, feel the structure of an unfamiliar space, and navigate toward solutions that could not have been specified before the navigation began.

This dual necessity is best illustrated by the shift from traditional «function specification» to what we might call «intent navigation». While traditional Computer Science education is built around Computational Thinking — where the student receives a precise specification, decomposes it, and constructs an algorithm — in the age of generative AI, this process is augmented by a prior phase of navigation.

For example, consider the task of building a personalized assistant for text analysis. In Phase 1 (Navigational Thinking), the learner does not know in advance which linguistic patterns the AI will identify. She enters an iterative dialogue — a practice of vibe-coding — using her tacit knowledge to probe the model's latent space. This results in the constitution of the problem space. In Phase 2 (Computational Thinking), once the intent is stabilized, the learner formalizes the solution by writing code, configuring APIs, and validating outputs. This two-phase structure demonstrates that navigation constitutes the problem space, while computation operates within it.

## 7. Objections and Responses

The argument developed in this paper will encounter predictable objections. Five of the most substantive are addressed here.

### Objection 1: Computational Thinking Is Sufficient

The first objection holds that Computational Thinking already provides everything that Computer Science education requires. The response turns on a structural observation about CT's presuppositions. Computational Thinking is designed for problems whose solution space can be fully specified before the solution process begins. The generative AI era has dramatically expanded a different class of problems: problems whose solution spaces must be constituted through navigation rather than specified before engagement begins. CT does not address this class of problems — not because it has failed, but because it was not designed for it. Navigational Thinking is not a replacement for CT; it is the epistemic mode required for the class of problems that precedes CT's application.

### Objection 2: Navigational Thinking Lacks Rigor

A second objection holds that navigational thinking is not a rigorous epistemic mode — it relies on intuition and tacit knowledge rather than precision and verifiability. This objection rests on a confusion between two different forms of rigor. The rigor of Computational Thinking is the rigor of explicit specification. But there is another form of rigor: the rigor of orientation in a geometrically structured space. The mathematician who solves an olympiad problem does not proceed vaguely — she moves with discipline and precision through an unfamiliar mathematical space, reads the local structure of the problem, and converges on a solution whose form was not known at the outset. Tacit knowledge, in Polanyi's analysis, is not the absence of knowledge but a different mode of knowing — structured, disciplined, and subject to correction.

### Objection 3: Navigational Thinking Is Just Trial and Error

A closely related objection holds that navigational thinking is simply trial-and-error search. The objection conflates two structurally different processes. Trial-and-error search is characterized by the independence of successive attempts: each trial is essentially unrelated to the previous ones, there is no accumulation of directional information, and the search space is sampled without geometric constraint. Navigational thinking is characterized by the geometric structure of successive steps. Each move is informed by what the previous move revealed about the local geometry of the space. The distinction can be made concrete: a person who proceeds by randomly trying different directions is engaged in trial-and-error. An experienced navigator navigating unfamiliar terrain reads the landscape, interprets the signals the environment offers, and moves with discipline toward a destination. Both may occasionally try an unproductive direction — but their processes are structurally different.

**Objection 4: Peirce and Papert Are Irrelevant to Computer Science**

A fourth objection questions the relevance of the philosophical frameworks mobilized in this paper. Neither is introduced as philosophical decoration. Peirce's analysis of indexical signification explains the operational mechanism of embedding spaces: why cosine similarity works, why positional relations encode semantic content, why the shift from symbolic to geometric representation is a change in the mode of signification. Without this analysis, the transition from CT to Navigational Thinking appears arbitrary. Papert's constructionism explains the pedagogical environment adequate to navigational knowledge production: why knowledge must be built through engagement with structured environments rather than transmitted through explicit instruction. Both are the analytical tools required to articulate what the geometric turn means for knowledge and for education.

**Objection 5: The Argument Will Be Obsolete When AI Architectures Change**

A final objection challenges the durability of the argument. Generative AI is a rapidly evolving field — the transformer architecture and specific dimensionalities discussed here may be superseded within a decade. This objection misidentifies the level at which the argument operates. The argument is not grounded in the specific architectural features of current generative AI systems. It is grounded in the mathematical properties of high-dimensional spaces as such: concentration of measure, near-orthogonality, exponential directional capacity, and manifold regularity. These properties are not features of particular architectures; they are features of high-dimensional geometry itself. Any system that operates in spaces of sufficient dimensionality will exhibit them, regardless of the specific computational mechanisms through which those spaces are instantiated. The argument is durable because it is grounded in mathematics, not in engineering.

## 8. Conclusion

This paper has argued that the geometric turn in generative AI is not merely a technical development. It is an epistemological event — one that transforms the conditions under which knowledge is produced, transmitted, and learned. The four structural properties of high-dimensional spaces examined in Section 2 — concentration of measure, near-orthogonality, exponential directional capacity, and manifold regularity — are not incidental features of a computational substrate. They are the conditions that make a new mode of knowledge production possible: navigational knowledge, grounded in indexical signification, enacted through structured traversal of learned manifolds.

The conceptual trajectory of the argument follows four connected movements. In low-dimensional spaces, meaning is grounded in distance: proximity indicates similarity, remoteness indicates difference. In high-dimensional spaces, this metric semantics collapses. Orientation becomes the primary carrier of meaning. The epistemological question migrates from «where?» to «whereto?» — from the location of a concept in a taxonomy to the direction it points in a relational field. This migration is not a metaphor; it is a precise description of the mathematical mechanism through which generative AI produces meaning.

The mode of signification has shifted correspondingly — from symbolic convention to indexical position. Peirce identified indexicality as a legitimate and important mode of meaning in the nineteenth century, but it remained a philosophical category without a technical realization at scale. Generative AI, operating through the geometry of high-dimensional embedding spaces, instantiates indexical signification as an operational mechanism at a scale previously unattainable. Embedding vectors signify not by arbitrary assignment but by occupying positions determined by the causal-statistical structure of the training process. The sign is really affected by its object — in Peirce's sense — and this real affection is what constitutes its meaning.

Papert's constructionism, developed in a symbolic era and constrained to symbolic microworlds, anticipated the navigational turn without being able to fully realize it. The generative AI system is the geometrized microworld that constructionism always required: high-dimensional, non-visualizable, governed by geometric constraints rather than explicit rules, navigated through continuous engagement rather than discrete command. The fundamental constructionist insight survives the transition intact: knowledge is built through engagement with structured environments, and the structure of the environment shapes the knowledge that can be constructed within it. What changes is the nature of the structure and the mode of engagement. Constructionism does not need to be abandoned in the age of generative AI. It needs to be geometrized.

These three shifts — from metric to directional, from symbolic to indexical, from programming to navigation — converge on a single educational consequence. Computational Thinking, first articulated by Papert himself in Mindstorms and subsequently developed into a foundational framework for Computer Science education, remains essential for the class of problems whose solution spaces are fully specifiable in advance. But Navigational Thinking and Computational Thinking are not simply complements: they are two sequential phases of the same epistemic process. While a problem remains indexical, navigation is operative. When the problem space stabilizes into symbolizable form, CT becomes applicable. Navigation constitutes the problem space; computation operates within it. The crisis of CT as a sole paradigm is not a failure of CT itself — it is a symptom of a structural shift in the nature of professional tasks. The class of tasks that remain indexical — tasks whose solution spaces must be navigated into existence — is growing rapidly. Vibe-coding is merely the visible tip of this iceberg. The iceberg is the new cognitive ecology of the generative AI era, in which Navigational Thinking and Computational Thinking must coexist as the two necessary phases of problem-solving.

What the age of generative AI demands is not the abandonment of rigor but its expansion: from the rigor of the algorithm to the rigor of the navigator, from the precision of the specification to the precision of the orientation.

Logic has not disappeared. It has been geometrized. And in that geometrization, a new educational epistemology — navigational, constructionist, indexical — comes into view. Understanding this epistemology is not an academic exercise. It is a precondition for preparing the next generation of thinkers for the spaces they will actually inhabit.